\documentclass[12pt,a4paper]{scrartcl}
\usepackage[utf8]{inputenc}
\usepackage[T1]{fontenc}
\usepackage{lmodern}
\usepackage[british]{babel}
\usepackage{amsmath}
\usepackage{amssymb}
\usepackage{amsfonts,ulem,bbm}
\usepackage{color}
\usepackage{float}
\usepackage{appendix}
\usepackage{graphicx}
\usepackage{epstopdf}
\usepackage{grffile}
\usepackage{hyperref}
\usepackage{textcomp}
\usepackage{subfigure,authblk}
\newcommand*\samethanks[1][\value{footnote}]{\footnotemark[#1]}
\newcommand{\E}{\ensuremath{\mathrm{e}}}
\newcommand{\I}{\ensuremath{\mathrm{i}\hspace{1pt}}}
\newcommand{\api}{\ensuremath{\text{a--}\pi}}
\newcommand{\aetap}{\ensuremath{\text{a--}\eta'}}
\newcommand{\afn}{\ensuremath{\text{a--}f_0}}
\newcommand{\SU}{\ensuremath{\mathrm{SU}}}
\newcommand{\arxiv}[2]{[arXiv:\,\href{http://arxiv.org/abs/#1}{\texttt{#1}} [\texttt{#2}]]}
\newcommand{\arxivold}[1]{[arXiv:\,\href{http://arxiv.org/abs/#1}{\texttt{#1}}\,]}
\title{
  {\vspace{-20mm}\normalsize
   \hfill\parbox[b][30mm][t]{35mm}{\textmd{MS-TP-18-04\\DESY 18-008}}}\\[-18mm]
The light bound states of $\boldmath{\mathcal{N}=1}$ supersymmetric SU(3)
Yang-Mills theory on the lattice}
\author[1]{Sajid Ali\thanks{\{sajid.ali,h.gerber,p.giudice,munsteg,scior\}@uni-muenster.de}}
\author[2,1]{Georg Bergner\thanks{georg.bergner@uni-jena.de}}
\author[1]{Henning Gerber\samethanks[1]}
\author[1]{Pietro Giudice\samethanks[1]}
\author[3]{Istvan Montvay\thanks{montvay@mail.desy.de}}
\author[1]{Gernot M\"unster\samethanks[1]}
\author[4]{Stefano Piemonte\thanks{stefano.piemonte@ur.de}}
\author[1]{Philipp Scior\samethanks[1]}
\affil[1]{University of M\"unster, Institute for Theoretical Physics,
Wilhelm-Klemm-Str.~9, D-48149 M\"unster, Germany}
\affil[2]{University of Jena, Institute for Theoretical Physics,
Max-Wien-Platz 1, D-07743 Jena, Germany}
\affil[3]{Deutsches Elektronen-Synchrotron DESY,
Notkestr.~85, D-22607 Hamburg, Germany}
\affil[4]{University of Regensburg, Institute for Theoretical Physics,
Universit\"atsstr.~31, D-93040 Regensburg, Germany}
\date{January 24, 2018}
\begin{document}
\maketitle
\begin{abstract}
In this article we summarise our results from numerical simulations of
$\mathcal{N}=1$ supersymmetric Yang-Mills theory with gauge group SU(3). We
use the formulation of Curci and Veneziano with clover-improved Wilson
fermions. The masses of various bound states have been obtained at different
values of the gluino mass and gauge coupling. Extrapolations to the limit of
vanishing gluino mass indicate that the bound states form mass-degenerate
supermultiplets.
\end{abstract}

\section{Introduction}

Supersymmetric gauge theories are building blocks for many extensions of the
Standard Model that aim to describe the physics of fundamental interactions
beyond the TeV scale. In general, the non-perturbative properties of these
theories play an essential role, in particular concerning the scenarios for a
dynamical supersymmetry (SUSY) breaking. Monte-Carlo simulations on the 
lattice are the method of choice for non-perturbative investigations of
quantum field theories. They provide a tool to investigate if and how the
theoretical predictions about supersymmetric theories are realised, see 
\cite{Bergner:2016sbv} for a more general discussion.

A specific application of numerical simulations of supersymmetric theories is
the determination of the spectrum of bound states and the study of the gluino
condensate in $\mathcal{N}=1$ supersymmetric Yang-Mills theory (SYM). In
fact, scenarios have been proposed for the multiplet structure of the
lightest bound states of this
theory~\cite{Veneziano:1982ah,Farrar:1997fn,Farrar:1998rm}, and several
analytical calculations have been presented for the chiral condensate,
see~\cite{Hollowood:1999qn} and references therein. The bound state spectrum
is the main focus of our current investigations. In our previous project we
have mainly considered SYM with gauge group SU(2). We have investigated the
particle spectrum of the theory and observed the expected degeneracy of the
states in the lightest supermultiplet~\cite{Bergner:2015adz}. We have now
extended our studies to the multiplet of excited states, and first results
have been reported in a contribution to the Lattice2017
conference~\cite{Ali:2017iof}.

Our most recent efforts are the numerical simulations of $\mathcal{N}=1$
supersymmetric Yang-Mills theory with gauge group SU(3). This theory is more
appealing from a phenomenological point of view, since it corresponds to the
gauge part of supersymmetric QCD. First results for this theory by our
collaboration can be found in~\cite{Feo:1999hw,Ali:2016zke,Ali:2017ijx}.
Another investigation of this theory is presented
in~\cite{Steinhauser:2017xqc}. Our simulations rely on the specific approach
of using Wilson fermions and a tuning of the gluino mass to restore chiral
symmetry and supersymmetry. As has been found by Veneziano and Curci, this
tuning is enough to recover both symmetries in the continuum
limit~\cite{Curci:1986sm}. We crosscheck the correctness of our tuning using
the supersymmetric Ward identities.

Simulations of theories with dynamical fermions in the adjoint representation
of SU(3), such as SYM, require significantly more resources than QCD with
quarks in the fundamental representation. Therefore, at present we are bound
to relatively small lattice sizes. The removal of the leading order lattice
cut-off terms from the fermion action is crucial in this case, since lattice
artefacts lead to an explicit supersymmetry breaking. From our previous
investigations we have found that the clover improved fermion action is
definitively a better choice than the unimproved stout smeared Wilson
fermions used in our first simulations~\cite{Ali:2016zke}. Further details of
our lattice formulation are explained in Section~\ref{sec:latticeact}.

The main focus of our project are the investigations of the lightest bound
states masses in SU(3) SYM. In particular, we want to check whether the mass
degeneracy between bosonic and fermionic particles expected in a
supersymmetric theory is realised non-perturbatively in the spectrum of bound
states. Analytic calculations based on low-energy effective actions predict a
supermultiplet of bound states consisting of mesonic gluinoballs and
fermionic gluino-glue particles~\cite{Veneziano:1982ah}. It was later
extended by an additional multiplet containg states created by glueball
operators~\cite{Farrar:1997fn,Farrar:1998rm}. We investigate the members of
both multiplets on the lattice by means of suitable operators. The results at
one value of the lattice spacing, presented in Section~\ref{sec:masses},
already indicate the expected formation of the supermultiplets of the
lowest-lying states as in the case of gauge group SU(2).

In Section~\ref{sec:scale} the methods for the determination of the
dimensionful reference scales are explained. Issues concerning the systematic
errors, like the check for an efficient sampling of topological sectors, are
discussed in Section~\ref{sec:syserr}. Finally, as an outlook we provide
first results at three additional lattice spacings in
Section~\ref{sec:contlim}.

\section{The improved lattice formulation of supersymmetric Yang-Mills
theory}
\label{sec:latticeact}

Supersymmetric SU(3) Yang-Mills theory describes gluons, the particles
associated with the non-Abelian gauge field for gauge group SU(3), and their
superpartners, the gluinos. Gluinos are Majorana fermions transforming under
the adjoint (octet) representation of SU(3). SU(3) SYM is of a complexity
comparable to QCD~\cite{Amati:1988ft}. It is expected that in the continuum
the particles described by this theory are bound states of gluons and
gluinos, that form supermultiplets degenerate in their masses, if
supersymmetry is unbroken. 
Since supersymmetry is broken explicitly by the lattice discretisation,
one important task of the project is to demonstrate that the data of the
numerical simulations are consistent with restoration of supersymmetry in the continuum limit.

In the continuum the (on-shell) Lagrangian of SYM, containing the gluon
fields $A_{\mu}$ and the gluino field $\lambda$, reads
\begin{equation}
\mathcal{L} = \mathrm{tr} \left[-\frac{1}{2}
F_{\mu\nu} F^{\mu\nu} + \I \bar{\lambda} \gamma^\mu
D_\mu \lambda -m_0 \bar{\lambda} \lambda \right] \,,
\end{equation}
where $F_{\mu\nu}$ is the non-Abelian field strength and $D_\mu$ denotes the
gauge covariant derivative in the adjoint representation. The gluino mass
term with the bare mass parameter $m_0$ breaks supersymmetry softly.

We employ the lattice formulation of SYM proposed by Curci and Veneziano
\cite{Curci:1986sm}. The gauge field is represented by link variables
$U_\mu(x)$. The corresponding gauge action is the Wilson action built from
the plaquette variables $U_p$. The gluinos are described by Wilson fermions
in the adjoint representation. In its basic form the lattice action reads
\begin{equation}
\mathcal{S}_L =
\beta \sum_p \left( 1 - \frac{1}{3} \mbox{Re}\,\mathrm{tr}\, U_p \right)
+ \frac{1}{2} \sum_{xy} \bar{\lambda}_x (D_w)_{xy} \lambda_y\,,
\end{equation}
with the Wilson-Dirac operator
\begin{eqnarray}
(D_w)_{x,a,\alpha;y,b,\beta}
    &=&\delta_{xy} \delta_{a,b} \delta_{\alpha,\beta}
    -\kappa \sum_{\mu=1}^{4}
      \left[ (1 - \gamma_\mu)_{\alpha,\beta}(V_\mu(x))_{ab}
                          \delta_{x+\mu,y} \right.\nonumber\\
    &&\left. + (1+\gamma_\mu)_{\alpha,\beta} (V^\dag_\mu(x-\mu))_{ab}
                          \delta_{x-\mu,y}\right],
\end{eqnarray}
where $V_{\mu}(x)$ are the link variables in the adjoint representation. The
hopping parameter $\kappa$ is related to the bare gluino mass via
$\kappa=1/(2m_0+8)$.

In our current simulations we have implemented the clover term in order to
reduce the leading lattice artefacts of the Wilson fermion action. The
additional term is
\begin{equation}
-   \frac{c_{sw}}{4}\, \bar{\lambda}(x) \sigma_{\mu\nu} F^{\mu\nu} \lambda(x).
\label{clover}
\end{equation}
where $F_{\mu\nu}$ is the clover plaquette. We have used the one-loop value
for the coefficient $c_{sw}$~\cite{Musberg:2013foa}, leading to a one-loop
$O(a)$ improved lattice action. This is a systematic and feasible approach
for setting the clover coefficient. Alternative tunings of the coefficient
are possible. In the SU(2) case we have tested a tadpole
resummation~\cite{Bergner:2015adz}, that leads to a considerable improvement of
the mass degeneracy for finite lattice spacings. 
At our current parameter range the value of $c_{sw}$ obtained with the proposed 
tadpole formula is not much different from the one-loop prediction.

The integration of the Majorana fermions yields
\begin{equation}
\int [d\lambda]\ \E^{- \frac{1}{2} \bar{\lambda} D_w \lambda}
= \mathrm{Pf}(C D_w) = \pm \sqrt{\det D_w}\,
\end{equation}
which is the Pfaffian of the Wilson-Dirac operator $D_w$ multiplied with the
charge conjugation matrix $C$. The square root of the determinant is handled
by the RHMC algorithm, whereas the sign of the Pfaffian has to be considered
in a reweighting of the observables. The effect of the Pfaffian sign is
discussed in Section~\ref{sec:pfaffsign}.

\section{Scale setting and simulation parameters}
\label{sec:scale}

We have performed simulations at four different values of the inverse gauge
coupling $\beta=5.4$, $5.5$, $5.6$, and $5.8$. The lattice size is
$16^3\times 32$, except for $\beta=5.4$, where we have chosen a $12^3\times
24$ lattice, and some additional large volume runs at $\beta=5.6$. The most
reliable results are obtained at $\beta=5.5$, whereas especially the results
at $\beta=5.8$ are in most cases excluded due to considerable finite size
effects, as discussed in detail in Section~\ref{sec:syserr}.

The scale is determined from an independent measurement of gluonic
observables in order to estimate the lattice spacing and the physical
volume. We are using two different quantities: the Sommer parameter $r_0$
and the scale $w_0$ from the gradient flow~\cite{Luscher:2010iy,Borsanyi:2012zs}. The
results in units of $r_0$ can be converted to QCD units fm or MeV using the
QCD scale setting $r_0=0.5$\,fm. The methods for the determination of
$r_0/a$ from a fit of the static quark-antiquark potential are explained in
our earlier work on SU(2) SYM~\cite{Bergner:2015adz}. 
At each $\beta$ the final values of the scales $w_0/a$ and $r_0/a$ are obtained by 
linearly extrapolating them as a function of the square of the adjoint pion mass to the chiral limit. 

The determination of $w_0/a$ follows the standard
methods~\cite{Luscher:2010iy,Borsanyi:2012zs} up to a modification of the
reference point. We have chosen a reference value of $u=0.2$ ($w_0^{0.2}$)
instead of the common value $0.3$ ($w_0^{0.3}$). This method is explained
in~\cite{Bergner:2014ska} and reduces the effect of topological freezing
that we observe at our smallest lattice spacings. The scaling between
$\beta=5.4$ and $\beta=5.5$ is compatible for $w_0^{0.2}/a$ and
$w_0^{0.3}/a$. Up to $\beta=5.6$ the scaling of $w_0^{0.2}$ is consistent
with the $r_0$ scaling. The smaller value of $u$ also considerably reduces
the quite large uncertainties for the chiral extrapolation of $w_0$ at
$\beta=5.6$.

\section{Signals for supersymmetry and chiral symmetry restoration}
\label{sec:symres}

As shown by Veneziano and Curci~\cite{Curci:1986sm}, supersymmetry and
chiral symmetry are restored in the continuum limit by the same tuning of
the bare gluino mass $m_0$. Chiral symmetry restoration cannot be probed
directly from the chiral Ward identities as in the case of two-flavour QCD,
since the U(1) axial symmetry is broken explicitly by an anomaly and not
only by the Wilson term. An alternative way is provided by the mass of the
adjoint pion ($\api$). The adjoint pion is not a physical particle in SYM.
It can, however, be defined by arguments based on the
OZI-approximation~\cite{Veneziano:1982ah}, or in the framework of partially
quenched chiral perturbation theory~\cite{Munster:2014cja}. When SYM is
considered as the partially quenched limit of Yang-Mills theory with two
Majorana flavours, the adjoint pion can be interpreted as a
pseudo-Nambu-Goldstone particle arising from spontaneous chiral symmetry
breaking. In the presence of an explicit chiral symmetry breaking by a
non-vanishing renormalised gluino mass, the square of the adjoint pion mass
scales proportional to the renormalised gluino mass. This relation is used
to extrapolate our numerical results to the chiral limit. The mass of the
adjoint pion ($m_{\pi}$) is measured from the exponential decay of the
connected part of the $\aetap$ meson correlator.

The reliability of our approach for the tuning to the chiral-supersymmetric
continuum limit has to be crosschecked with different prescriptions.
Supersymmetric Ward identities provide an alternative solid signal for the
remnant chiral symmetry breaking without further assumptions about the
structure of chiral effective actions. Another approach is to determine the
transition point for the discrete subgroup of chiral symmetry that is left
unbroken by the anomaly. Below we discuss in detail the theoretical
expectations and the results of these different tunings towards the chiral
limit. All of these signals must agree in the continuum limit, but we will
show that already at finite lattice spacings the discrepancies are small and
even negligible compared to other systematic uncertainties.

We have investigated the supersymmetric Ward identities at many different
values of our bare mass parameter. A combination of the supercurrent
renormalisation constant ($Z_S$) and the renormalised gluino mass ($m_S$)
can be determined from this measurement. The techniques of the measurement
and analysis can be found in~\cite{Farchioni:2001wx,Ali:2018}. We developed
a generalised least squares method~\cite{Ali:2017nug,Ali:2018} to obtain
more reliable estimates of $am_SZ^{-1}_S$ and its statistical error.
Note that the tuning of the clover coefficient in the
fermion action up to one-loop order does not ensure automatically the $O(a)$
improvements of the SUSY Ward identities, as explained
in~\cite{Farchioni:2001wx}. Further perturbative calculations of improvement
coefficients would be required to reach an $O(a)$ scaling of the same order,
and we plan to investigate this aspect in the future.

\begin{figure}[thb]
 \begin{center}
   \includegraphics[width=0.6\textwidth,clip]{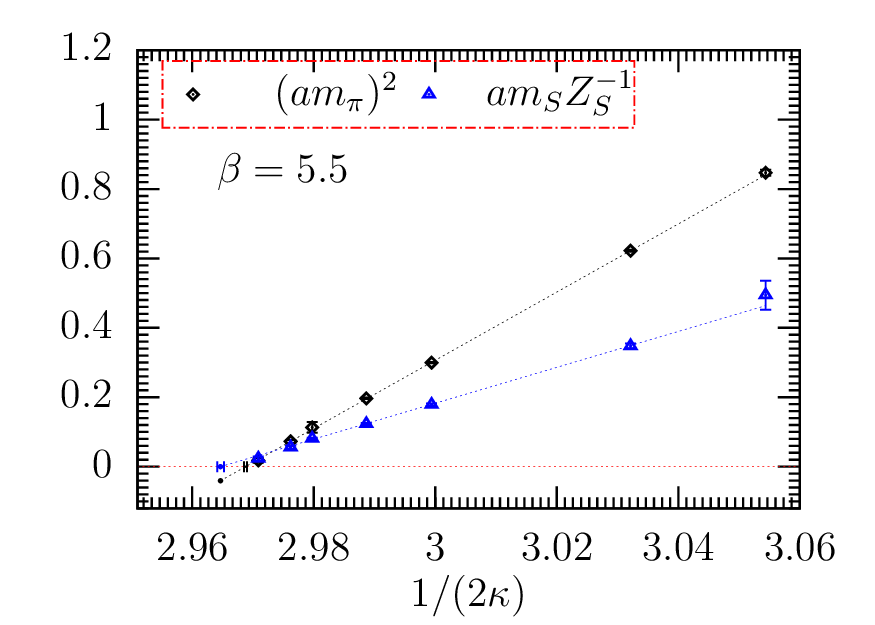}
 \end{center}
\caption{\label{KcGCS}Linear extrapolations of the adjoint pion mass
squared ($m^2_{\pi}$) (black) and the renormalised gluino mass
($am_SZ^{-1}_S$) obtained by SUSY Ward identities (blue) as a function of
the bare parameter $\kappa$ towards the chiral point ($\kappa_c$).}
\end{figure}

The value of the critical parameter $\kappa_c$, where the renormalised
gluino mass vanishes, is obtained from an extrapolation of $m_{\pi}^2$ to
zero, and it is compared with the determination from the supersymmetric Ward
identities, as shown in Figure~\ref{KcGCS}. The two values of $\kappa_c$ are
very close to each other, but there is a small difference of around
$0.00023(5)$. This discrepancy is presumably due to lattice artifacts, and
is expected to disappear in the continuum limit. Results at other values of
$\beta$ in view of the continuum limit are discussed in \cite{Ali:2018}.

\begin{figure}[thb]
 \begin{center}
  \subfigure[Single Gaussian fit]{\includegraphics[width=0.45\textwidth]{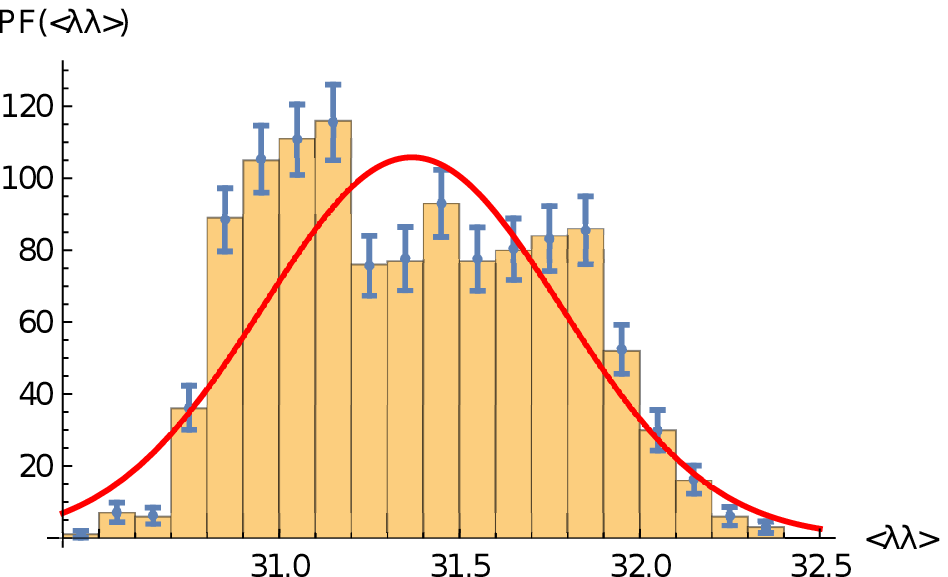}}
  \subfigure[Double Gaussian fit]{\includegraphics[width=0.45\textwidth]{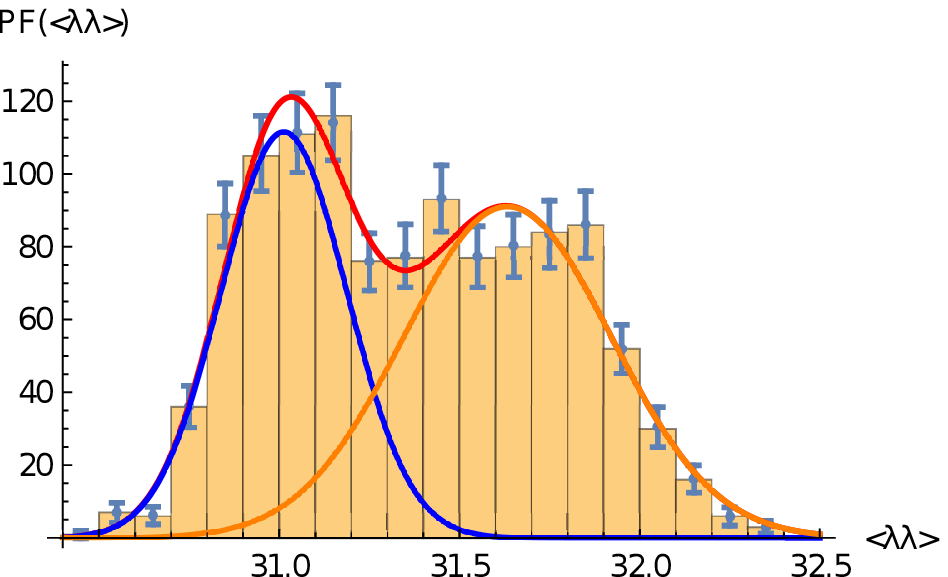}}
\caption{Double peak structure of the histogram of the chiral condensate
at $\kappa=0.1665$ and $\beta=5.6$ from 1200 configurations. A single
Gaussian does not fit our data, which are instead consistent with a sum of
two Gaussian functions.}
\label{fig:gluino_double}
  \end{center}
\end{figure}

A third determination of $\kappa_c$ could be found by studying the change of
the gluino condensate at zero temperature occurring at the chiral phase
transition. In fact, chiral symmetry is the invariance of the continuum
action of SYM with respect to the U(1) rotation of the fermion field
\begin{equation}
 \lambda \rightarrow \exp{\{-\I \theta \gamma_5\}} \lambda\,.
\end{equation}
An anomaly breaks chiral symmetry down to $Z_{2Nc}$ at the quantum level,
such that only 2$N_c$ values of $\theta$ leave the partition function
invariant. At zero temperature even the discrete group $Z_{2Nc}$ is broken
down spontaneously to $Z_2$ by a non-vanishing expectation value of the
gluino condensate. The coexistence of $N_c$ degenerate vacua is a signal for
a first order phase transition crossing the chiral limit as a function of
the gluino mass. We can search the transition in the chiral condensate
$\langle\bar{\lambda}\lambda\rangle$, corresponding to the real part of the gluino
condensate. The condensate on each configuration should hence fluctuate
between two distinct values at the critical point $\kappa_c$. However,
simulations close to the critical point are very difficult and we are
limited to small volumes, like $6^4$ and $8^4$, to ensure the convergence of
the inverter of the Dirac-Wilson operator. We have used periodic boundary
conditions to reduce the breaking of supersymmetry that would otherwise
appear at non-zero temperature due to the difference between the thermal
statistics of fermions and bosons. Running our simulations at $\beta=5.6$,
we find signals of a double-peak structure of the chiral condensate at
$\kappa=0.1665$, see Figure~\ref{fig:gluino_double}. The determined value of
$\kappa_c$ is consistent with the other determinations
($\kappa_c(m_\pi)=0.16635(4)$,
$\kappa_c(\langle\bar{\lambda}\lambda\rangle)=0.1662(4)$). Taking into
account the uncertainties from finite size effects, we can conclude that
there is a good agreement with the theoretical expectations concerning the
vacuum structure of the theory at zero temperature.

\section{Results for the lightest supermultiplet in supersymmetric SU(3)
Yang-Mills theory}
\label{sec:masses}

The estimation of the masses of bound states in SU(3) SYM is the main focus
of the current work. 
The particle spectrum is composed of bound states of gluons and gluinos. 
One expects composites of gluino fields (gluinoballs), of gluon fields (glueballs), 
and of both (gluino-glueballs). The physical states would be mixtures of those.
The measurement of these particles is quite challenging. For the mesonic gluinoballs, which 
are flavour singlet mesons, and the glueballs this can be understood by comparison with 
the corresponding particles in QCD.
Hence a rather large statistics is required
in order to get reliable estimates for the masses. As detailed in
Section~\ref{sec:symres}, the adjoint pion mass is used for the
extrapolations of the masses to the chiral limit. All masses are determined
from the exponential decay of the correlators in the corresponding channels.
Further details of the different measurements are explained in the
following. First we consider the simulations at $\beta=5.5$ on a $16^3\times
32$ lattice, since these are our most precise and most reliable results.

\subsection{Glueballs, gluino-glueballs and the supermultiplet formation}
\begin{figure}[htb]
  \begin{center}
  \subfigure[Glueball]{\includegraphics[width=0.45\textwidth]{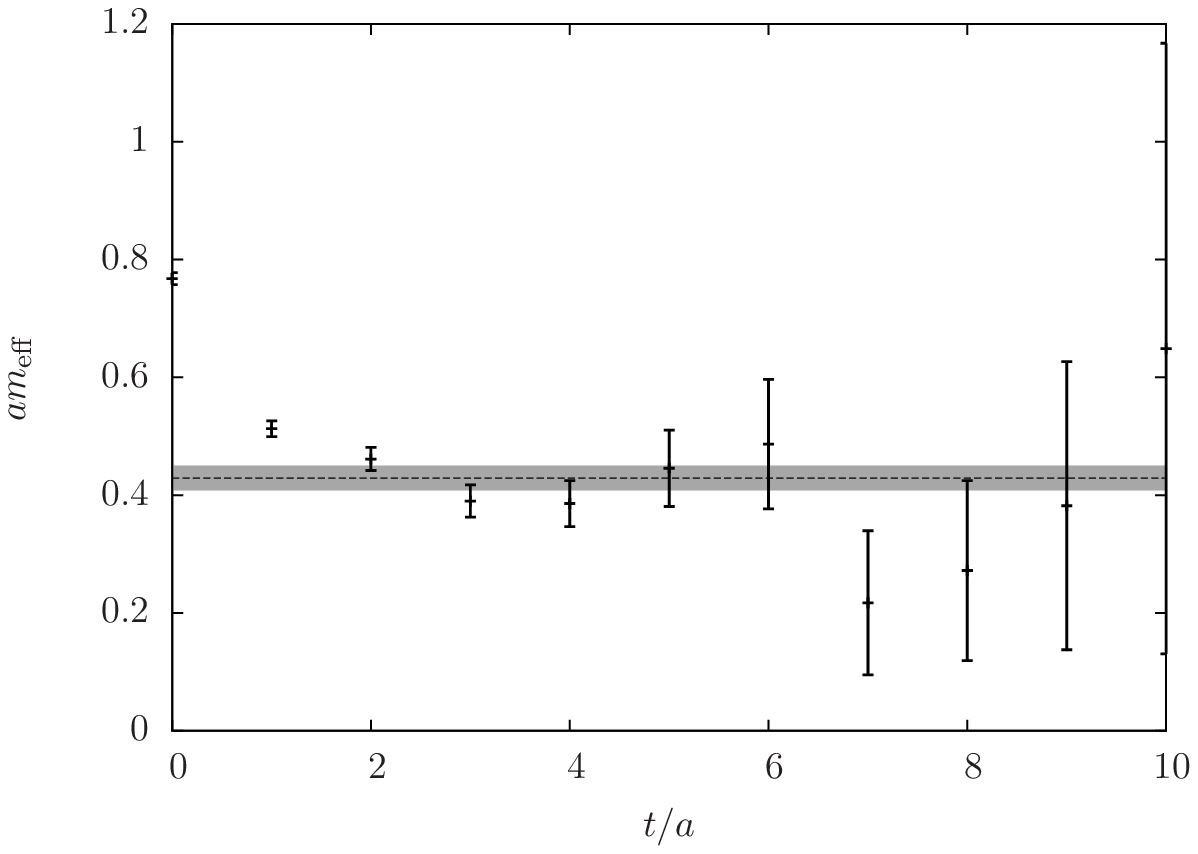}\label{fig:glueballeffmass}}
  \subfigure[Gluino-glue]{\includegraphics[width=0.45\textwidth]{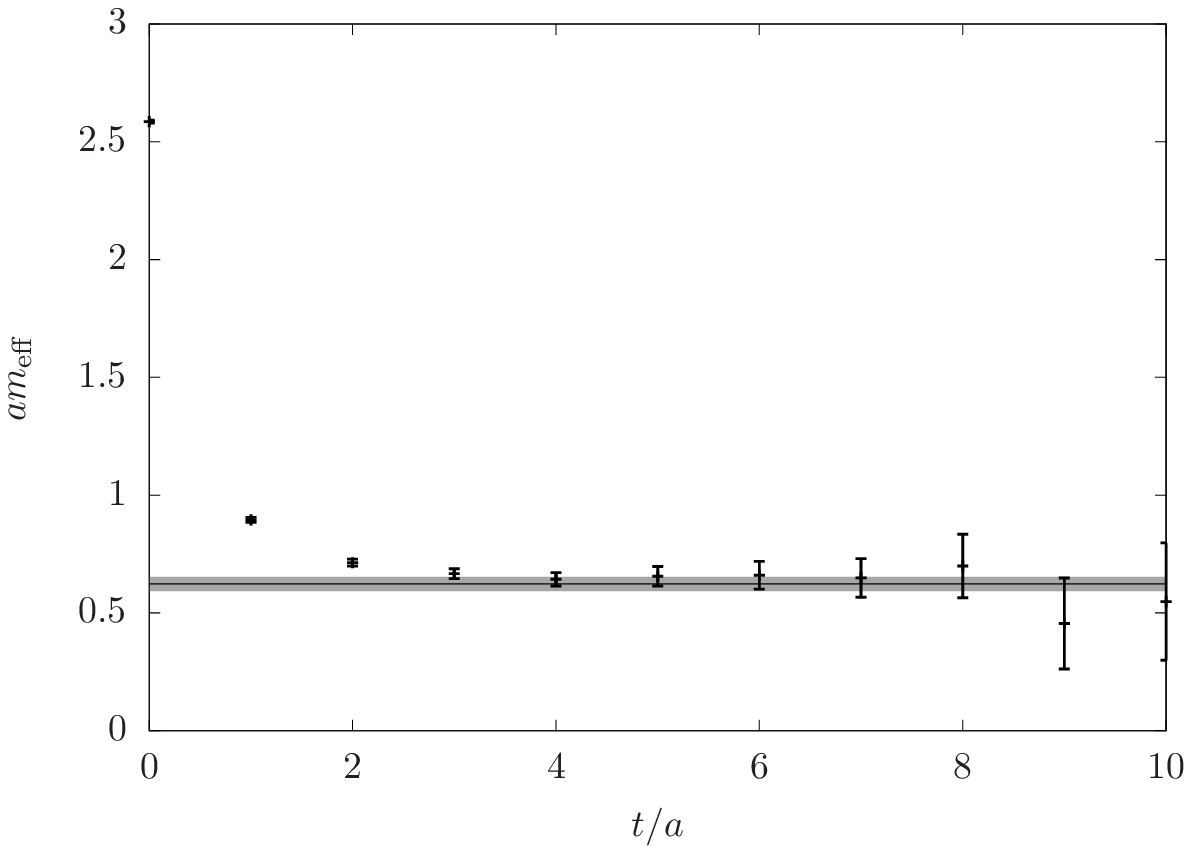}\label{fig:ggeffmass}}
\end{center}
 \caption{Mass plateau of the gluino-glue and the $0^{++}$ glueball for
$\beta=5.5$, $\kappa=0.1673$, on a $16^3\times 32$ lattice. The final value
indicated by the gray line is obtained from a fit of the correlation
function.}
 \end{figure}
\begin{figure}[htb]
\centering
\includegraphics[width=0.65\textwidth]{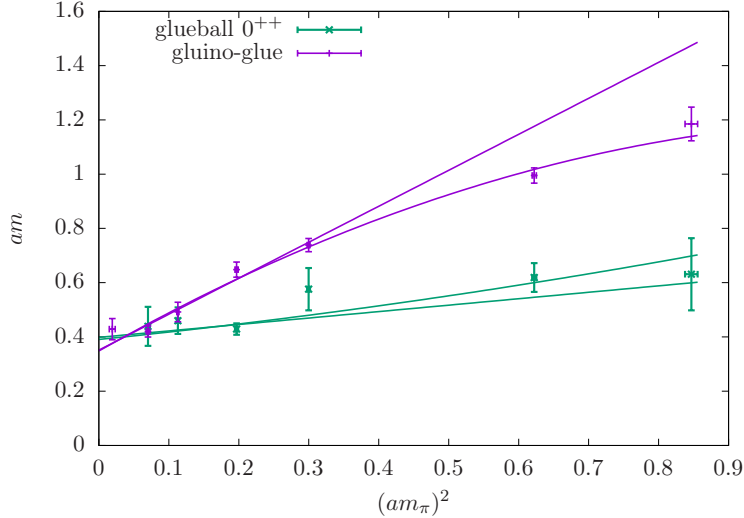}
\caption{\label{plot3} The masses of the gluino-glue and the $0^{++}$
glueball at $\beta=5.5$ in lattice units. The figure shows a linear and a
quadratic fit of the data. The two largest values of the adjoint pion mass
$m_\pi$ are excluded in the linear fit of the gluino-glue.}
\end{figure}

A reliable determination of the glueball masses is challenging. In the
current work we have focused on the $0^{++}$ glueball, since it provides
the best signal-to-noise ratio. We have applied variational methods to
reduce the effects of excited states. 
More about our methods is explained in our work
on SU(2) SYM~\cite{Bergner:2015adz} and in~\cite{Ali2:2018}. An example of an effective mass
determined from the exponential decay of the correlators is shown in
Figure~\ref{fig:glueballeffmass}.

The fermionic partners of the glueballs are particles created from gluino
and gluon fields. The corresponding operator is composed of the field
strength $F_{\mu\nu}$ and the gluino field,
\begin{equation}
\label{eq:gg}
\tilde{O}_{g\tilde{g}}=
\sum_{\mu\nu} \sigma_{\mu\nu} \textrm{Tr} \left[ F^{\mu\nu} \lambda\right],
\end{equation}
with $\sigma_{\mu\nu}=\frac{1}{2} \left[ \gamma_\mu,\gamma_\nu \right]$.
$F_{\mu\nu}$ is represented by the clover plaquette on the lattice. The
measurement of this particle from the ensemble of gauge configurations uses
Jacobi and APE smearing techniques in order to improve the signal and to
suppress the excited state contaminations. Figure~\ref{fig:ggeffmass} shows
an example of the effective mass and of the quality of our fits. Further
details can be found in our earlier publications~\cite{Bergner:2015adz}.

As in our previous investigations of SU(2) SYM, the degeneracy of the
fermionic gluino-glue and its bosonic partners is the most important signal
for the supermultiplet formation. The chiral extrapolation of the $0^{++}$
glueball and gluino-glue masses at $\beta=5.5$ are shown in
Figure~\ref{plot3}. At large adjoint pion masses, corresponding to a large
soft supersymmetry breaking, the gluino-glue is about twice as heavy as the
glueball. However, in the chiral limit the two masses become degenerate up
to our current statistical precision.

In general, supersymmetry breaking lattice artefacts, indicated by a mass gap between the states 
of the lightest supermultiplet, are expected at any finite lattice spacing.
In our current simulations at $\beta=5.5$ their influence seems to be under control since
the mass gap is not significantly larger than other uncertainties of the measurements. 
At these parameters our simulations therefore are already close enough to the
continuum limit to reproduce main features of the continuum
theory.

\subsection{The completion of the chiral multiplet by mesonic gluinoballs}

\begin{figure}[thb]
 \subfigure[$\aetap$]{\includegraphics[width=0.45\textwidth]{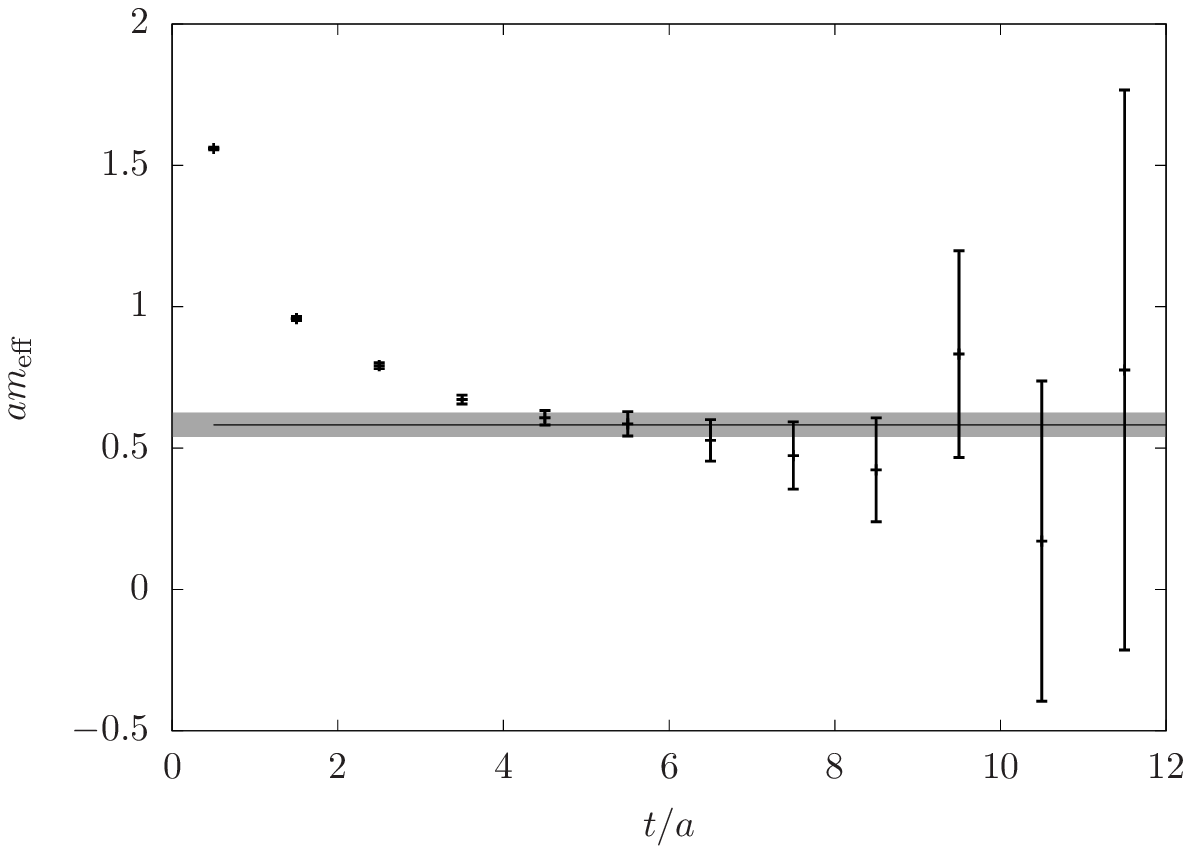}}
 \subfigure[$\afn$]{\includegraphics[width=0.45\textwidth]{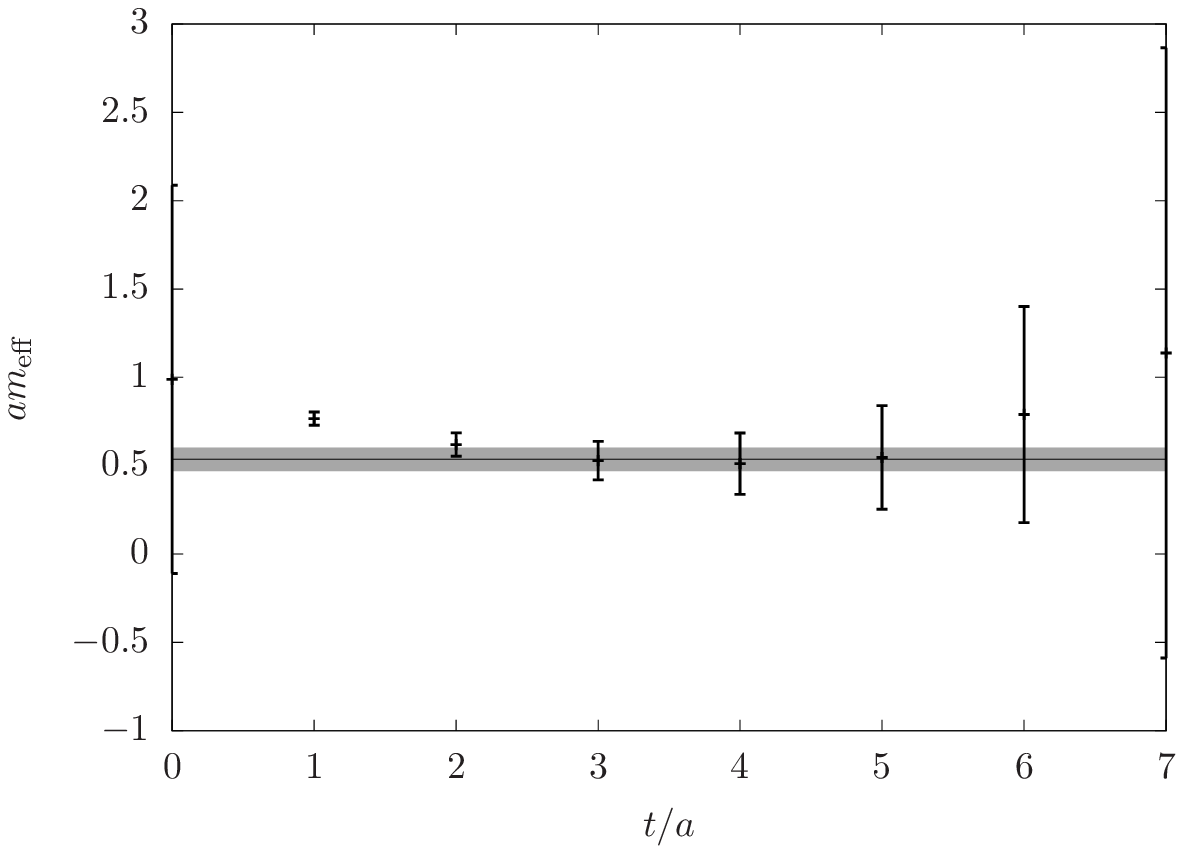}}
\caption{Mass plateau for the mesons $\aetap$ and $\afn$ for $\beta=5.5$,
$\kappa=0.1673$, on a $16^3\times 32$ lattice. The final value indicated by
the gray line is obtained from a fit of the correlation function.
 \label{fig:meson}}
\end{figure}
\begin{figure}[thb]
\centering
\includegraphics[width=0.65\textwidth]{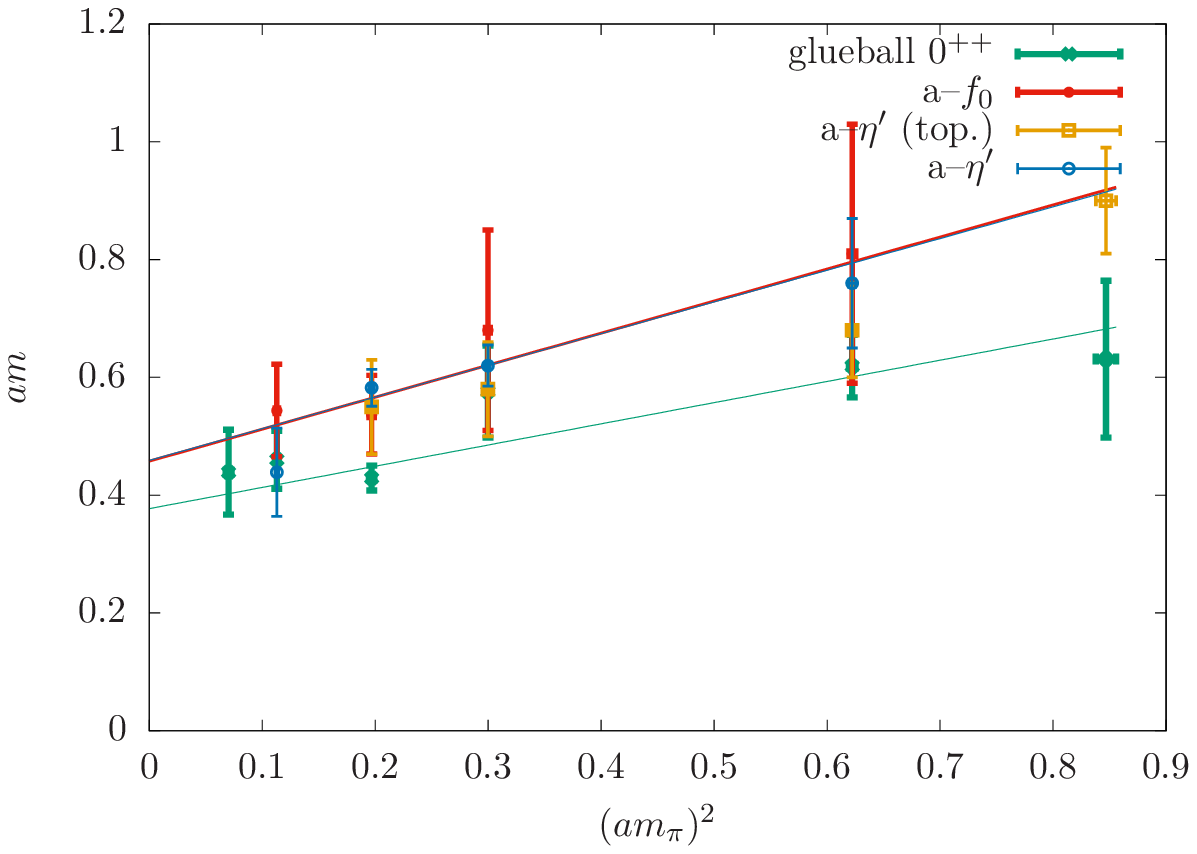}
\caption{\label{fig:bosons} The masses of the different bosonic particles
(singlet scalar and pseudoscalar mesons) in comparison to the glueball mass
at $\beta=5.5$.
$\aetap$(top.) indicates the pseudoscalar meson mass obtained from the
topological charge density correlator. }
\end{figure}

Two chiral supermultiplets are expected to represent the degrees of freedom
at low energies. The supermultiplet of the lightest bound states should
contain a scalar, a pseudoscalar, and a fermionic particle. The scalar and pseudoscalar 
particles described by the low-energy effective 
actions are either of glueball or of mesonic type. The actual states would, however, 
be mixtures of those, and on the lattice the states created by the corresponding 
glueball and mesonic operators cannot be distinguished unambiguously.

In our previous investigation of SU(2) SYM we have found that
the mesonic operator provides a better signal for the lightest pseudoscalar
state, whereas the scalar meson is degenerate with the scalar glueball. We
complete the results for the masses of the lowest multiplet with the
additional data from the $\aetap$ meson for the pseudoscalar channel and use
the $\afn$ as a cross check for the scalar glueball data.

The singlet mesonic operators are named similar to their QCD counterparts,
the pseudoscalar $\aetap$ ($\bar{\lambda}\gamma_5\lambda$) and the scalar
$\afn$ ($\bar{\lambda}\lambda$). The disconnected part is an essential
contribution to the correlation functions of these particles. The signal for
this part of the correlators is rather noisy. Our methods for the
measurement include truncated eigenmode approximation and preconditioning to
improve the signal, see~\cite{Bergner:2011zz} for further details. The
results contain still quite large uncertainties, see Figure~\ref{fig:meson},
but we are able to obtain the first estimates of the masses also in the
mesonic channel. We have also done an alternative determination of the
$\aetap$ mass from the topological charge density
correlator~\cite{Fukaya:2015ara}, which is in good agreement with the
results obtained from the mesonic correlators.

The signal in the scalar and pseudoscalar channels can be improved using a
variational approach combining different mesonic and gluonic operators. We
have recently tested this in SU(2) SYM~\cite{Ali:2017iof}, and we plan to
use it also in the SU(3) case. Most likely it will reduce the remnant
excited state contamination of the ground state signal, especially in the
mesonic sector.

As shown in Figure~\ref{fig:bosons}, the masses of the scalar and
pseudoscalar mesons are almost degenerate with the scalar glueball mass
within our current precision. Similar to the scalar glueball, they show only
a weak fermion mass dependence. There is indeed a formation of a complete
supersymmetry multiplet in the chiral limit and the $\afn$ provides a signal
for the same lightest scalar state as the glueball.

The final extrapolated values of the particle masses in units of the Sommer
scale $r_0$ are\nopagebreak
\begin{center}
\begin{tabular}{|l|l|l|l|}
\hline
  gluino-glue & glueball $0^{++}$ & $\aetap$ &  $\afn$  \\
\hline
  2.83(44) & 3.22(95) & 3.70(71) &  3.69(63)  \\
\hline
\end{tabular}
\end{center}

\section{Estimation of systematic uncertainties}
\label{sec:syserr}

Several systematic uncertainties of our current results need to be
considered. The most important limitation is the remnant supersymmetry
breaking by the lattice discretisation. We are currently not able to perform
a complete extrapolation to the continuum, but as explained in
Section~\ref{sec:contlim}, the remaining uncertainties are currently at the
order of the statistical errors. We plan more in-depth investigations to get
a better signal for the continuum extrapolation. The finest accessible
lattice spacing is limited by two effects. Since the number of lattice
points is limited, the finer lattice spacing leads to a smaller volume. The
lattice spacing can, consequently, only be reduced until the finite volume
effects significantly affect the results. The second important effect is the
topological freezing, that leads to large autocorrelation times when the
lattice spacing becomes too small. As we will discuss below, our simulations
at $\beta=5.6$ and $\beta=5.8$ are affected by these effects.

\subsection{The Pfaffian sign}
\label{sec:pfaffsign}
\begin{figure}
\centering
\includegraphics[width=0.65\textwidth]{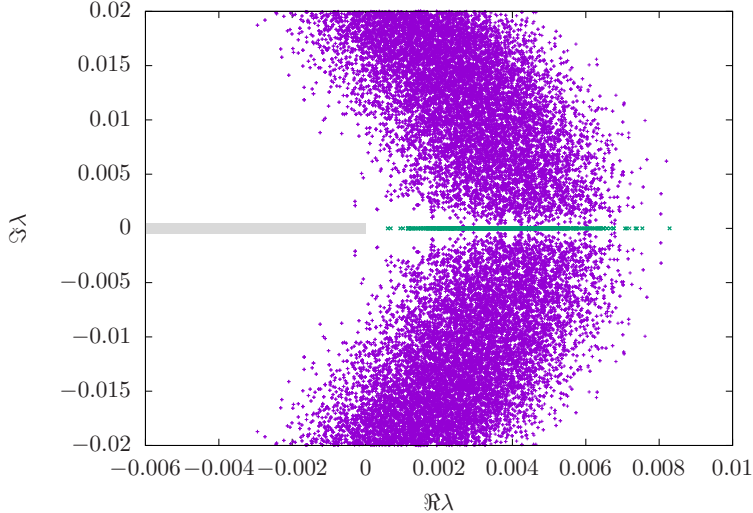}
\caption{\label{fig:pfaffian} The eigenvalues of the Dirac-Wilson operator
in the complex plane, relevant for the sign of the Pfaffian, from 1022
configurations at $\beta=5.5$, $\kappa=0.1683$. The green points indicate
eigenvalues with a significant chirality of the eigenvector $\langle v |
\gamma_5 | v \rangle > 0.001$. Possible Pfaffian sign changes might occur if
there are real negative eigenvalues, i.~e.\ green points in the shaded
region.}
\end{figure}
The sign of the Pfaffian has to be taken into account in our simulations.
We expect that the Pfaffian sign is not significant at our
current parameters, but this has to be confirmed by measurements. The
fluctuations of the Pfaffian sign become more significant at smaller gluino
masses and coarser lattices. It is hence possible to approach the chiral
continuum limit from simulations without a relevant contribution of the
sign. However, we have to check explicitly that we are indeed in the region
without relevant negative sign contributions. It is enough to consider the
run with the smallest gluino mass to confirm this. As explained in our
earlier investigations, the sign of the Pfaffian is obtained from the number
of degenerate pairs of real negative eigenvalues of the Dirac-Wilson
operator, see also~\cite{Bergner:2011zp} for the methods of this
measurement. Sign changes are hence only possible, if there are negative
real eigenvalues. As shown in Figure~\ref{fig:pfaffian} we do not observe
any of these negative eigenvalues at least for a large subset of
configurations. We conclude that the Pfaffian sign for the current runs at
$\beta=5.5$ is not relevant.

\subsection{Finite size effects}
\begin{figure}
\centering
\includegraphics[width=0.65\textwidth]{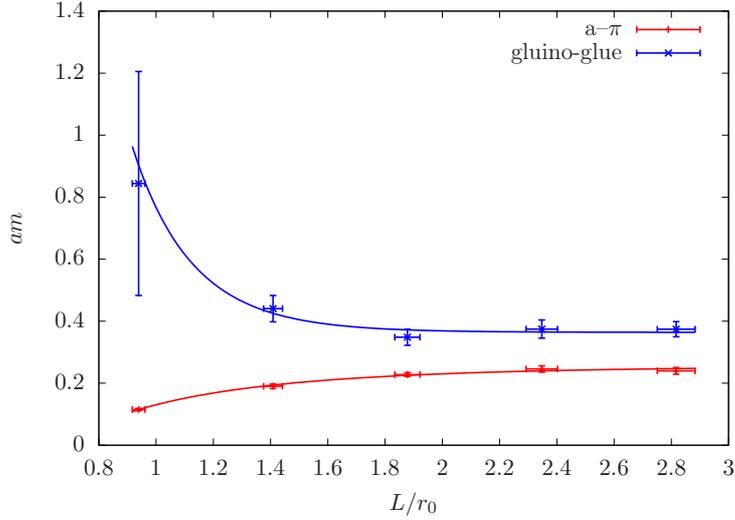}
\caption{\label{fig:finitesize} The adjoint pion and gluino-glue mass from
simulations at $\beta=5.6$, $\kappa=0.1660$ on lattices of size $8^3\times
32$, $12^3\times 32$, $16^3\times 32$, $20^3\times 40$, and $24^3\times 48$.
These data indicate the magnitude of finite size effects. The results have
been fitted assuming a finite size correction proportional to $\exp(-\alpha
L)/L$ with some positive coefficient $\alpha$.}
\end{figure}

Finite size effects play an important role in the estimations of the mass
spectrum. In earlier investigations with gauge group SU(2) we have found
that for small volumes the gluino-glue gets heavier and the degeneracy of
the spectrum is lost, but larger volumes ($L/r_0>2.4$) are not
affected~\cite{Bergner:2012rv}. We check whether similar finite volume
effects also appear for gauge group SU(3) at a rather small lattice spacing
($\beta=5.6$). The results shown in Figure~\ref{fig:finitesize} have a
considerable uncertainty for the gluino-glue data. At the lattice size of
$N_s=16$, where we have performed most of the simulations, the adjoint pion
mass shows around 10\% finite size effects. We conclude that on a
$16^3\times 32$ lattice the finite size effects are negligible at
$\beta=5.5$; at $\beta=5.6$ they are of the order of our current still quite
limited accuracy. At $\beta=5.8$ rather large finite size effects are
expected. Consequently we have focussed here on $\beta=5.5$ and plan to
increase the lattice volume in our future more precise simulations at
$\beta=5.6$.

\begin{figure}
\begin{center}
\subfigure[$\beta$ dependence]{\includegraphics[width=0.45\hsize]{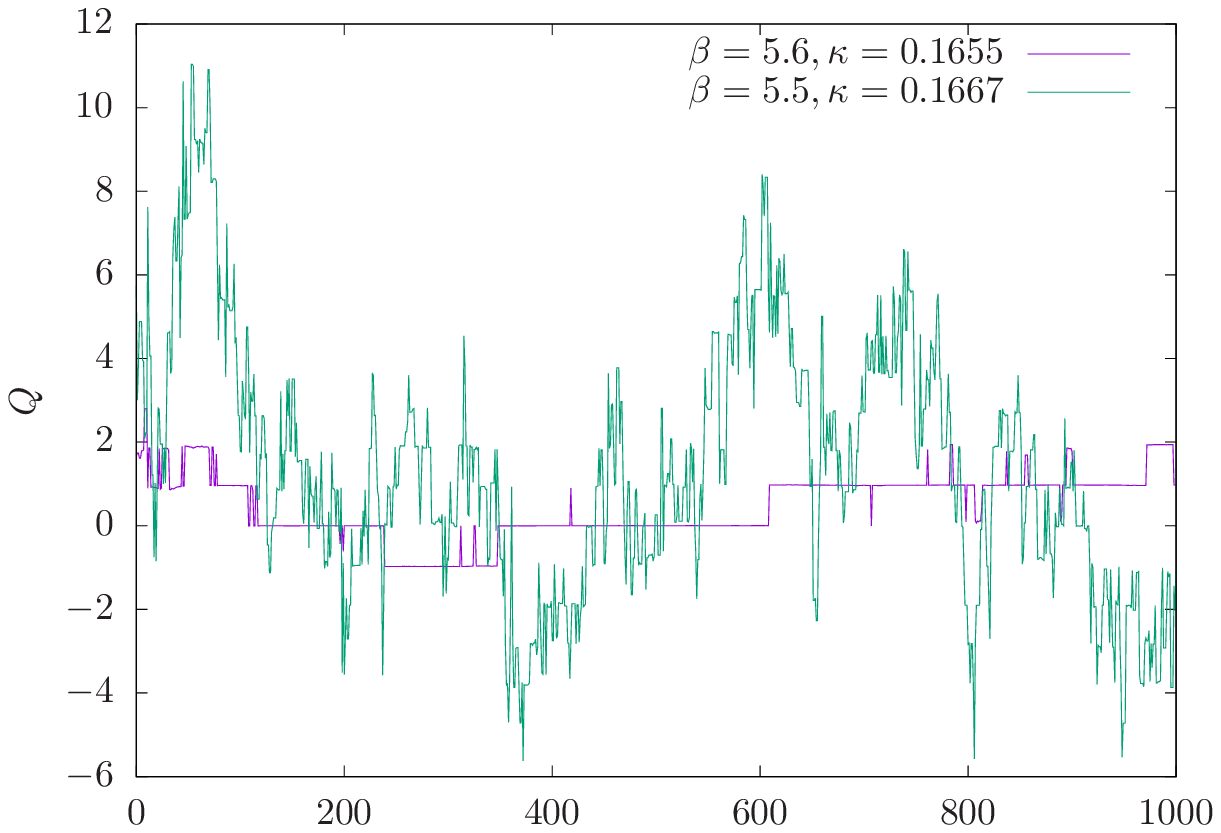}}
\subfigure[volume dependence]{\includegraphics[width=0.45\hsize]{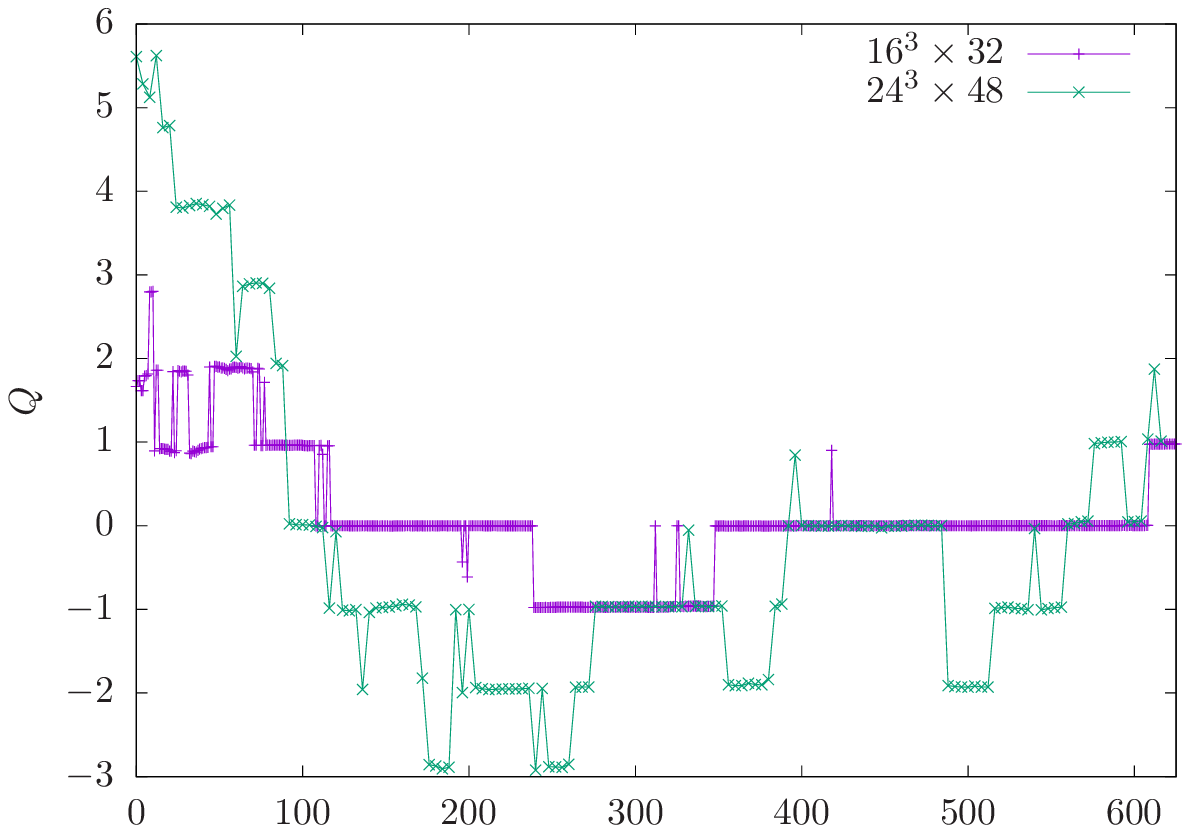}}
\end{center}
\caption{\label{fig:tophist} Left hand side: The history of the topological
charge from simulations at $\beta=5.5$ and $\beta=5.6$ on a $16^3\times 32$
lattice. Right hand side: The history of the topological charge at
$\beta=5.6$, $\kappa=0.1660$ for two different volumes.}
\end{figure}

\subsection{The sampling of topological sectors}
As known from QCD, topological sectors are not efficiently sampled at
lattice spacings smaller than roughly $0.05$\,fm, leading to the loss of ergodicity of
Monte-Carlo lattice simulations. Very large autocorrelation times are
especially observed for topological quantities, and the topological charge
is effectively frozen towards the continuum limit. Our simulations are
already at a very fine lattice spacing, therefore we must ensure a
reasonable sampling of the topological sectors. We have measured the average
topological charge $\langle Q \rangle$, the corresponding integrated
autocorrelation time $\tau_Q$, and the topological susceptibility $\chi_Q$.
The topological freezing is under control for $\beta = 5.5$ ($\tau_Q$ is
between 17 and 46), but starts to become more significant at $\beta = 5.6$
($\tau_Q$ is between 58 and 185), which might also be related to the small
volume, see Figure~\ref{fig:tophist}. The histogram of the topological
charge for $\beta=5.8$ shows a nearly frozen topology. Hence, for this
$\beta$ the values of $\tau_Q$ and other quantities are not reliable.

\section{Outlook: continuum limit and comparison to the SU(2) case}
\label{sec:contlim}
\begin{figure}[ht]
\centering
\includegraphics[width=0.65\textwidth]{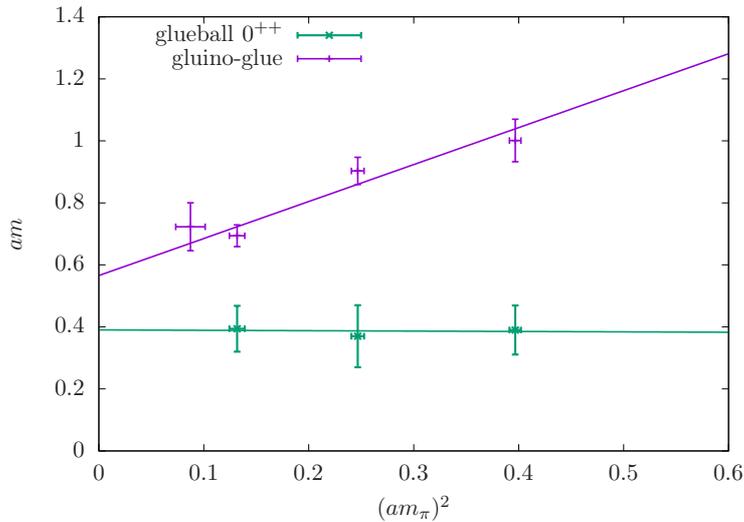}
\caption{\label{fig:beta54gg} The masses of the fermionic gluino-glue and
the bosonic $0^{++}$ glueball at ${\beta=5.4}$ in lattice units. 
The figure includes an extrapolation of the chiral limit based on a linear fit.}
\end{figure}
Currently our most precise results are from $\beta=5.5$. The results from
the finer lattices are limited by finite size effects and topological
freezing. In addition we have done simulations at one coarser lattice at
$\beta=5.4$. Our first preliminary results for the glueball and the
gluino-glue at the coarse lattice spacing are shown in
Figure~\ref{fig:beta54gg}. There is a considerable gap between the
constituents of the multiplet, but further investigations are required to
crosscheck the chiral extrapolations.

\begin{figure}[ht]
\centering
\subfigure[scaling with $w_0^{0.2}$]{\includegraphics[width=0.45\textwidth]{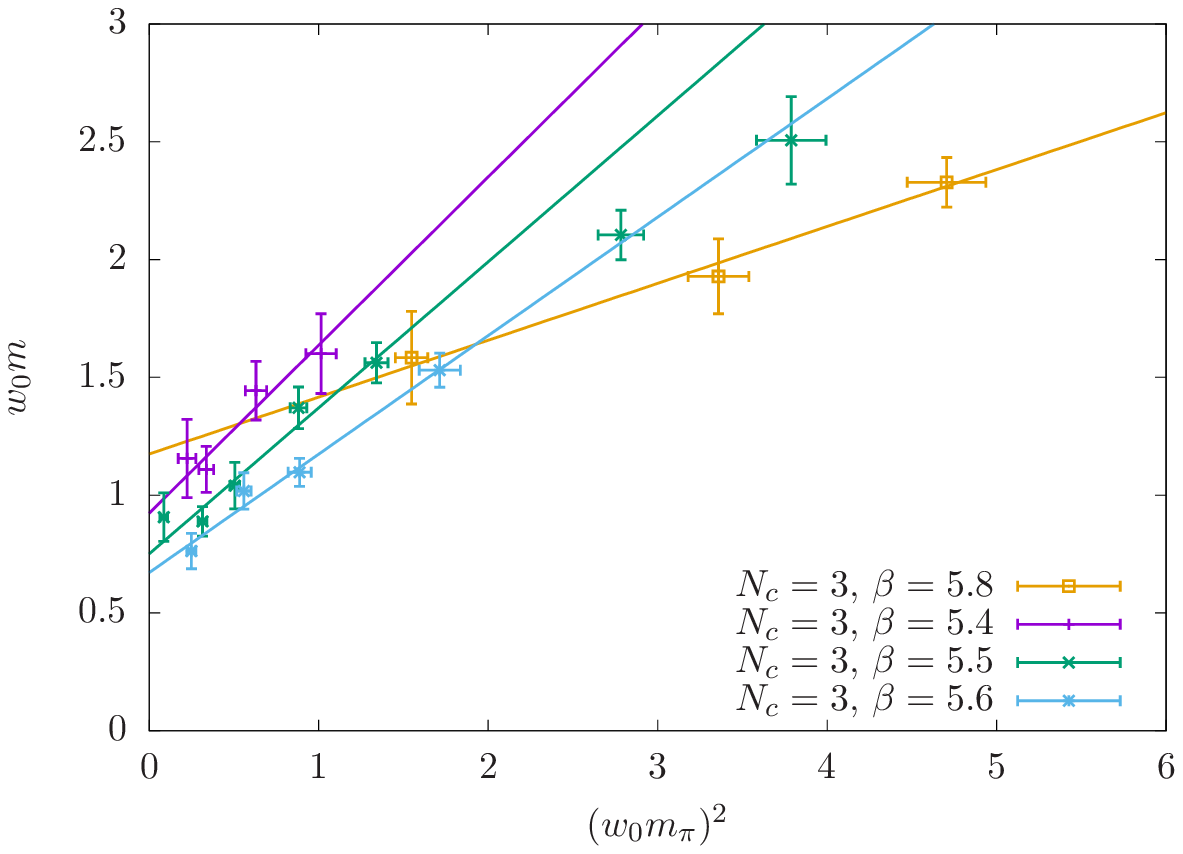}}
\subfigure[scaling with $r_0$]{\includegraphics[width=0.45\textwidth]{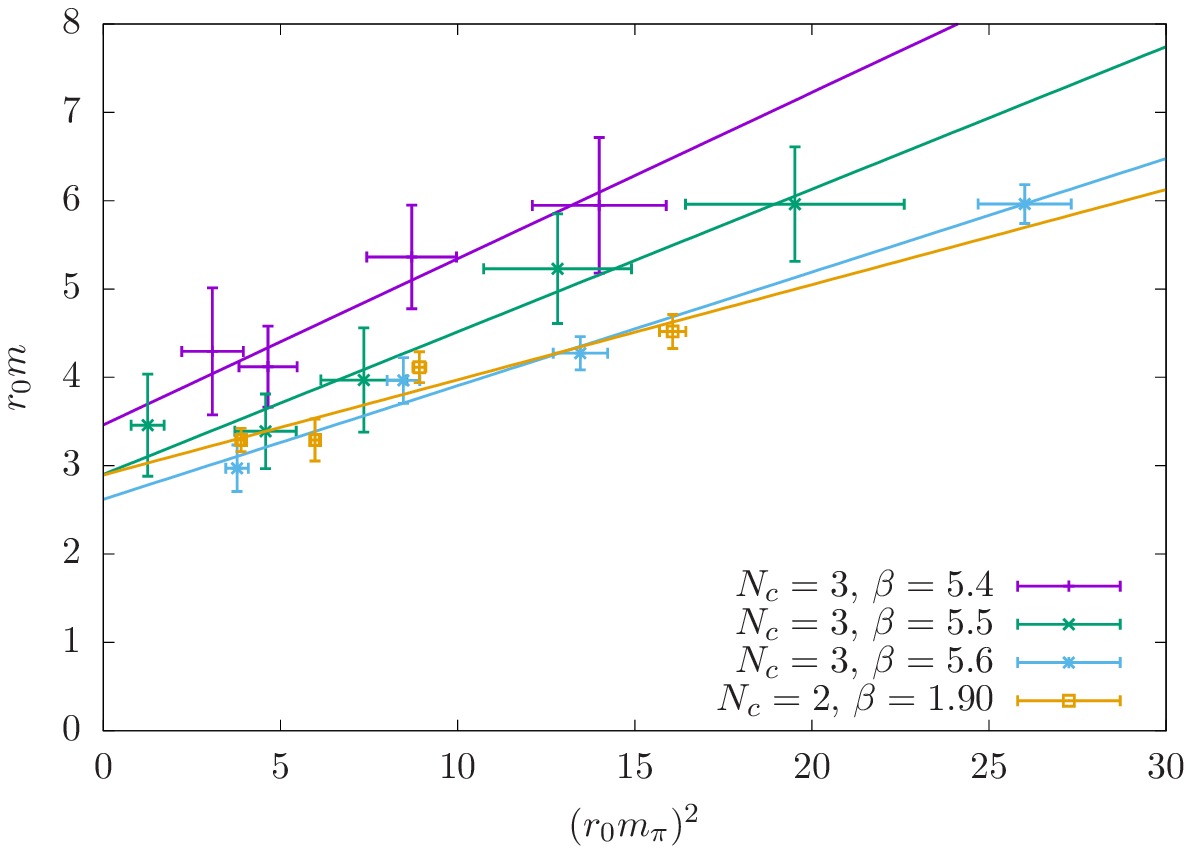}}
\caption{\label{fig:scalinggg} A comparison of the gluino-glue mass in units
of the Sommer scale $r_0$ and of $w_0^{0.2}$ for different lattice spacings
(corresponding to different $\beta$) and $\SU(N_c)$ gauge groups as a
function of the squared adjoint pion mass. The simulations of SU(3) SYM at $\beta=5.4$,
$5.5$, $5.6$, and $5.8$ have been done with a clover improved
fermion action and a plain Wilson gauge action; the
simulations of SU(2) SYM at $\beta=1.9$ with a plain Wilson fermion action, one
level of stout smearing, and a tree level Symanzik improved gauge action.}
\end{figure}
With the current results we are already able to make the first estimations
of the scaling towards the continuum limit. The results in units of the
scales $w_0^{0.2}$ and $r_0$ for the gluino-glue are shown in
Figure~\ref{fig:scalinggg}. As expected, the results at $\beta=5.8$ are
unreliable. From the other lattice spacings a trend towards smaller
gluino-glue masses is seen in the continuum limit.

The preliminary results of the extrapolations at different $\beta$ in units
of the scale $w_0^{0.2}$ are\nopagebreak
\begin{center}
\begin{tabular}{|l|l|l|}
\hline
$\beta$ &  gluino-glue & glueball $0^{++}$ \\
\hline
5.4 & 0.90(13) & 0.6240(59)  \\
5.5 & 0.743(77) & 0.84(20)  \\
5.6 & 0.673(66) & 0.60(15)  \\
\hline
\end{tabular}
\end{center}

The masses in units of $r_0$ can be compared to our previous results for
SU(2) SYM as shown in Figure~\ref{fig:scalinggg}. It is remarkable that the
masses of the lightest states in physical units are comparable for SU(2) and
SU(3) SYM.

\section{Conclusions}
\label{sec:conclusions}

Our current results for the lightest bound states in supersymmetric SU(3)
Yang-Mills theory on the lattice already provide strong indications for a
supermultiplet formation in the chiral continuum limit. As in the case of
gauge group SU(2), there is no unexpected signature for supersymmetry
breaking, and the lattice artefacts can be controlled. The non-perturbative
numerical investigation of this theory with Wilson fermions is hence
possible and the theoretical suggestions of Veneziano and Curci can be
applied in practice.

The simulation parameters are quite restricted, since an SU(3) Yang-Mills
theory with fermions in the adjoint representation demands considerably more
computational efforts than in the case of the fundamental representation.
The implementation of an improved fermion action is hence quite essential,
as shown by our numerical results.

Further studies at larger lattices are required to complete the continuum
extrapolation and reduce finite volume uncertainties. Nevertheless, the
present data already indicate that the multiplet formation persists towards 
the continuum limit. We are currently also exploring more advanced methods to reduce the
excited state contamination from the determination of the masses of the
bound spectrum.

\section{Acknowledgements}

The authors gratefully acknowledge the Gauss Centre for Supercomputing
e.~V.\,\linebreak(www.gauss-centre.eu) for funding this project by providing
computing time on the GCS Supercomputer JUQUEEN and JURECA at J\"ulich
Supercomputing Centre (JSC) and SuperMUC at Leibniz Supercomputing Centre
(LRZ). Further computing time has been provided on the compute cluster PALMA
of the University of M\"unster. This work is supported by the Deutsche
Forschungsgemeinschaft (DFG) through the Research Training Group ``GRK 2149: 
Strong and Weak Interactions - from Hadrons to Dark Matter''.
G.B. acknowledges support from the
Deutsche Forschungsgemeinschaft (DFG) Grant No. BE 5942/2-1.


\appendix 
\section{Data}
\label{sec:data}

In the following we list the data of our most relevant ensembles. The main
part of the paper considers the runs on a $16^3\times 32$ lattice at
$\beta=5.5$ with a one-loop clover coefficient $c_{sw}=1.598$. These are
summarised in the following table:
\begin{center}
\begin{small}
\begin{tabular}{|c|c|c|c|c|c|c|c|c|}
\hline
 $\kappa$ &$am_{\pi}$ &$r_0/a$ &$am_{gg}$ &$am_{gpp}$ &$am_{\eta}$ &$am_{f_0}$ &$am_{f_0,\text{top}}$ &$N_\text{configs}$ \\
\hline
0.1637 &0.9202(49)&3.86(11)&1.185(62)&0.63(13)&--&--&0.900(90)&500 \\
\hline
0.1649 &0.7888(19)&4.39(16)&0.995(28)&0.619(53)&0.76(11)&0.81(22)&0.680(80)&3212 \\
\hline
0.1667 &0.5475(19)&5.56(13)&0.739(24)&0.576(78)&0.620(35)&0.68(17)&0.580(80)&4415 \\
\hline
0.1673 &0.4437(26)&6.15(16)&0.648(28)&0.429(21)&0.582(31)&0.537(67)&0.550(80)&5984 \\
\hline
0.1678 &0.3360(23)&6.88(33)&0.492(36)&0.460(49)&0.439(75)&0.544(79)&--&3591 \\
\hline
0.168 &0.2651(51)&8.31(40)&0.420(21)&0.439(72)&--&--&--&2673 \\
\hline
0.1683 &0.138(15)&8.96(50)&0.429(39)&--&--&--&--&1645 \\
\hline
\end{tabular}
\end{small}
\end{center}

All the quantities are in units of the lattice spacing $a$. The summarised
quantities are the Sommer parameter $r_0/a$ and the masses of the $\api$
($am_{\pi}$), the gluino-glue ($am_{gg}$), the $0^{++}$ glueball
($am_{gpp}$), the $\aetap$ meson ($am_{\eta}$), and the $\afn$ meson
($am_{f_0}$). In addition there is an alternative measurement of the $\afn$
meson mass from the correlator of the topological charge density
($am_{f_0,\text{top}}$). $N_\text{configs}$ is the number of generated thermalised
configurations. Currently not the complete statistic is used in all
measurements.

\end{document}